\documentclass[iop]{emulateapj}
\usepackage{graphicx}

\slugcomment{To appear in ApJ.}

\begin{document}

\title{Investigating the TeV Morphology of MGRO J1908+06 with VERITAS}

\shortauthors{Aliu et al.}

\author{
E.~Aliu\altaffilmark{1},
S.~Archambault\altaffilmark{2},
T.~Aune\altaffilmark{3},
B.~Behera\altaffilmark{4},
M.~Beilicke\altaffilmark{5},
W.~Benbow\altaffilmark{6},
K.~Berger\altaffilmark{7},
R.~Bird\altaffilmark{8},
J.~H.~Buckley\altaffilmark{5},
V.~Bugaev\altaffilmark{5},
J.~V~Cardenzana\altaffilmark{9},
M.~Cerruti\altaffilmark{6},
X.~Chen\altaffilmark{10,4},
L.~Ciupik\altaffilmark{11},
E.~Collins-Hughes\altaffilmark{8},
M.~P.~Connolly\altaffilmark{12},
W.~Cui\altaffilmark{13},
J.~Dumm\altaffilmark{14},
V.~V.~Dwarkadas\altaffilmark{15},
M.~Errando\altaffilmark{1},
A.~Falcone\altaffilmark{16},
S.~Federici\altaffilmark{4,10},
Q.~Feng\altaffilmark{13},
J.~P.~Finley\altaffilmark{13},
H.~Fleischhack\altaffilmark{4},
P.~Fortin\altaffilmark{6},
L.~Fortson\altaffilmark{14},
A.~Furniss\altaffilmark{17},
N.~Galante\altaffilmark{6},
D.~Gall\altaffilmark{18},
G.~H.~Gillanders\altaffilmark{12},
S.~Griffin\altaffilmark{2},
S.~T.~Griffiths\altaffilmark{18},
J.~Grube\altaffilmark{11},
G.~Gyuk\altaffilmark{11},
D.~Hanna\altaffilmark{2},
J.~Holder\altaffilmark{7},
G.~Hughes\altaffilmark{4},
T.~B.~Humensky\altaffilmark{19},
P.~Kaaret\altaffilmark{18},
M.~Kertzman\altaffilmark{20},
Y.~Khassen\altaffilmark{8},
D.~Kieda\altaffilmark{21},
F.~Krennrich\altaffilmark{9},
S.~Kumar\altaffilmark{7},
M.~J.~Lang\altaffilmark{12},
A.~S~Madhavan\altaffilmark{9},
G.~Maier\altaffilmark{4},
A.~J.~McCann\altaffilmark{22},
K.~Meagher\altaffilmark{23},
J.~Millis\altaffilmark{24},
P.~Moriarty\altaffilmark{25},
R.~Mukherjee\altaffilmark{1},
D.~Nieto\altaffilmark{19},
A.~O'Faol\'{a}in de Bhr\'{o}ithe\altaffilmark{8},
R.~A.~Ong\altaffilmark{3},
A.~N.~Otte\altaffilmark{22},
D.~Pandel\altaffilmark{26},
N.~Park\altaffilmark{27},
M.~Pohl\altaffilmark{10,4},
A.~Popkow\altaffilmark{3},
H.~Prokoph\altaffilmark{4},
J.~Quinn\altaffilmark{8},
K.~Ragan\altaffilmark{2},
J.~Rajotte\altaffilmark{2},
G.~Ratliff\altaffilmark{11},
L.~C.~Reyes\altaffilmark{28},
P.~T.~Reynolds\altaffilmark{29},
G.~T.~Richards\altaffilmark{23},
E.~Roache\altaffilmark{6},
J.~Rousselle\altaffilmark{3},
G.~H.~Sembroski\altaffilmark{13},
K.~Shahinyan\altaffilmark{14},
F.~Sheidaei\altaffilmark{21},
A.~W.~Smith\altaffilmark{21},
D.~Staszak\altaffilmark{2},
I.~Telezhinsky\altaffilmark{10,4},
K.~Tsurusaki\altaffilmark{18},
J.~V.~Tucci\altaffilmark{13},
J.~Tyler\altaffilmark{2},
A.~Varlotta\altaffilmark{13},
V.~V.~Vassiliev\altaffilmark{3},
S.~Vincent\altaffilmark{4},
S.~P.~Wakely\altaffilmark{27},
J.~E.~Ward\altaffilmark{5},
A.~Weinstein\altaffilmark{9},
R.~Welsing\altaffilmark{4},
A.~Wilhelm\altaffilmark{10,4}
}

\altaffiltext{1}{Department of Physics and Astronomy, Barnard College, Columbia University, NY 10027, USA}
\altaffiltext{2}{Physics Department, McGill University, Montreal, QC H3A 2T8, Canada}
\altaffiltext{3}{Department of Physics and Astronomy, University of California, Los Angeles, CA 90095, USA}
\altaffiltext{4}{DESY, Platanenallee 6, 15738 Zeuthen, Germany}
\altaffiltext{5}{Department of Physics, Washington University, St. Louis, MO 63130, USA}
\altaffiltext{6}{Fred Lawrence Whipple Observatory, Harvard-Smithsonian Center for Astrophysics, Amado, AZ 85645, USA}
\altaffiltext{7}{Department of Physics and Astronomy and the Bartol Research Institute, University of Delaware, Newark, DE 19716, USA}
\altaffiltext{8}{School of Physics, University College Dublin, Belfield, Dublin 4, Ireland}
\altaffiltext{9}{Department of Physics and Astronomy, Iowa State University, Ames, IA 50011, USA}
\altaffiltext{10}{Institute of Physics and Astronomy, University of Potsdam, 14476 Potsdam-Golm, Germany}
\altaffiltext{11}{Astronomy Department, Adler Planetarium and Astronomy Museum, Chicago, IL 60605, USA}
\altaffiltext{12}{School of Physics, National University of Ireland Galway, University Road, Galway, Ireland}
\altaffiltext{13}{Department of Physics, Purdue University, West Lafayette, IN 47907, USA }
\altaffiltext{14}{School of Physics and Astronomy, University of Minnesota, Minneapolis, MN 55455, USA}
\altaffiltext{15}{Department of Astronomy and Astrophysics, University of Chicago, Chicago, IL, 60637}
\altaffiltext{16}{Department of Astronomy and Astrophysics, 525 Davey Lab, Pennsylvania State University, University Park, PA 16802, USA}
\altaffiltext{17}{Santa Cruz Institute for Particle Physics and Department of Physics, University of California, Santa Cruz, CA 95064, USA}
\altaffiltext{18}{Department of Physics and Astronomy, University of Iowa, Van Allen Hall, Iowa City, IA 52242, USA}
\altaffiltext{19}{Physics Department, Columbia University, New York, NY 10027, USA}
\altaffiltext{20}{Department of Physics and Astronomy, DePauw University, Greencastle, IN 46135-0037, USA}
\altaffiltext{21}{Department of Physics and Astronomy, University of Utah, Salt Lake City, UT 84112, USA}
\altaffiltext{22}{Kavli Institute for Cosmological Physics, University of Chicago, Chicago, IL 60637, USA}
\altaffiltext{23}{School of Physics and Center for Relativistic Astrophysics, Georgia Institute of Technology, 837 State Street NW, Atlanta, GA 30332-0430}
\altaffiltext{24}{Department of Physics, Anderson University, 1100 East 5th Street, Anderson, IN 46012}
\altaffiltext{25}{Department of Life and Physical Sciences, Galway-Mayo Institute of Technology, Dublin Road, Galway, Ireland}
\altaffiltext{26}{Department of Physics, Grand Valley State University, Allendale, MI 49401, USA}
\altaffiltext{27}{Enrico Fermi Institute, University of Chicago, Chicago, IL 60637, USA}
\altaffiltext{28}{Physics Department, California Polytechnic State University, San Luis Obispo, CA 94307, USA}
\altaffiltext{29}{Department of Applied Physics and Instrumentation, Cork Institute of Technology, Bishopstown, Cork, Ireland}


\begin{abstract}

We report on deep observations of the extended TeV gamma-ray source MGRO J1908+06 made with the VERITAS very high energy (VHE) gamma-ray observatory.  Previously, the TeV emission has been attributed to the pulsar wind nebula (PWN) of the {\it Fermi}-LAT pulsar PSR J1907+0602.  We detect MGRO J1908+06 at a significance level of 14 standard deviations (14~$\sigma$) and measure a photon index of $2.20\pm0.10_{\rm stat}\pm0.20_{\rm sys}$.  The TeV emission is extended, covering the region near PSR J1907+0602 and also extending towards SNR G40.5--0.5.  When fitted with a 2-dimensional Gaussian, the intrinsic extension has a standard deviation of $\sigma_{src} = 0.44^\circ\pm0.02^\circ$.  In contrast to other TeV PWNe of similar age in which the TeV spectrum softens with distance from the pulsar, the TeV spectrum measured near the pulsar location is consistent with that measured at a position near the rim of G40.5--0.5, $0.33\arcdeg$ away.

\end{abstract}


\section{Introduction}\label{sec:Introduction}

MGRO J1908+06 was discovered by the Milagro collaboration during their seven-year survey of the Northern Hemisphere in very high energy (VHE; $E>100$ GeV) gamma rays at a median energy of 20 TeV \citep{Abdo07}.  Observations with the High Energy Stereoscopic System (H.E.S.S.) of the location of MGRO J1908+06 were performed as part of a Galactic plane survey \citep{hess_plane}, and after the announcement of the Milagro source, follow-up observations confirmed an extended VHE source, HESS J1908+063 \citep{Aharonian09}.  The extended VHE source was also confirmed by observations with VERITAS \citep [Very Energetic Radiation Imaging Telescope Array System;][]{ward}.  These results noted the possibility of an association with the nearby shell-type supernova remnant (SNR) G40.5-0.5 \citep{green09}.  However, the measured extension of the VHE source, well beyond the boundary of the radio SNR, required either an additional source of gamma rays or the presence of dense molecular matter with which ultra-relativistic particles could interact and produce gamma rays \citep{Aharonian09}.  \citet{Abdo09} discovered a radio-quiet gamma-ray pulsar, PSR J1907+0602, in the vicinity of MGRO J1908+06 with the Large Area Telescope (LAT) on the {\it Fermi} satellite.   A later investigation by the {\it Fermi}-LAT team concluded that MGRO J1908+06 and HESS J1908+063 were the pulsar wind nebula (PWN) of PSR J1907+0602 \citep{Abdo10b}.  (Hereafter, J1908 will be used to generically refer to the VHE source.)

Milagro observed emission from J1908 with a flux corresponding to $\sim$80\% of the Crab Nebula flux at 20~TeV and an upper limit on the intrinsic source extension of $2.6^\circ$ \citep{Abdo07}.  The lower energy threshold and better angular resolution of H.E.S.S.\ allowed further investigation of the morphology of the source.  A fit of a 2D-Gaussian function to the excess map gave a centroid of RA=$286.98^{\circ}$ and Dec=$6.27^{\circ}$ ($l=40^{\circ}23'9''.2\pm2'.4_{\rm stat}$ and $b=-0^{\circ}47'10''.1\pm2'.4_{\rm stat}$ with systematic errors of $20\arcsec$ per axis) and an intrinsic extension of $\sigma_{\rm src}=0.34^{\circ}~^{+0.04}_{-0.03}$ \citep{Aharonian09}.

PSR J1907+0602 has a characteristic age of 19.5~kyr and a spin-down power of $3\times10^{36}$ erg/s \citep{Abdo10b}.  The time-averaged flux density at 1.51~GHz was 3.4~$\mu$Jy in measurements made with the Arecibo 305~m radio telescope.  From the position and the dispersion measure, the distance was estimated to be 3.2~kpc with a nominal error of $20\%$ \citep{Abdo10b}.  Chandra observations of the location of PSR J1907+0602 revealed a faint X-ray counterpart with a non-thermal component that is possibly extended \citep{Abdo10b}. 

\citet{Downes80} estimated the age of SNR G40.5-0.5 to be 20-40 kyr, and its distance is estimated by \citet{yang} to be 3.4 kpc.  Both quantities are compatible with the measured properties of PSR J1907+0602; thus the two objects may be physically associated.  However, distance estimates using other methods imply a distance of 5.5--8.5 kpc \citep{yang,Downes80} for the SNR.  The {\it Fermi}-LAT team used a Very Large Array Galactic Plane Survey \citep[VGPS;][]{Stil} 1420 MHz continuum image to estimate the position of the SNR to be RA=$286.79^\circ$, Dec=$6.50^\circ$, with an angular diameter of $0.43^\circ$. At a distance  of 3.2~kpc, the separation between the SNR and the pulsar is then 28~pc \citep{Abdo10b}.  If the pulsar is the progenitor of SNR G40.5-0.5, and assuming that the pulsar originated at the center of the SNR, with the characteristic age of 19.5 kyr, the transverse velocity of the pulsar would be 1400 km/s \citep{Abdo10b}.  This velocity is $\sim$3 times higher than the typical velocity given to a pulsar due to a supernova ``kick" \citep{slanereview}, but three other pulsars are estimated to have transverse velocities above 1000 km/s:  PSR B2224+65 \citep[$\sim1600$~km/s; ][]{pulsar1}, PSR B2011+38 \citep[$\sim1600$~km/s; ][and references therein]{pulsar2}, and PSR B1508+55 \citep[$\sim1100$~km/s; ][]{pulsar3}.  Table \ref{tab:sources} summarizes the parameters of PSR J1907+0602 and SNR G40.5--0.5.

VERITAS observed J1908 repeatedly during 2007-2012, obtaining a total exposure of 62~hours.  Here, we describe this deep TeV observation taken in order to help understand the physical origin of the emission.  We describe the observations in Section 2 and analysis in Section 3.  We present our results on the morphology and spectrum of the TeV emission in Section 4.  We discuss interpretations of the results in Section 5.

\begin{deluxetable}{lcc}
\tablecaption{Parameters of PSR J1907+0602 and SNR G40.5--0.5 
  \label{tab:sources}}
\tablehead{
Parameter       & PSR J1907+0602  & SNR G40.5--0.5}
\startdata
Right Ascension & $286.978^\circ$ & 286.786$^\circ$ \\ 
Declination     & $6.038^\circ$   & 6.498$^\circ$   \\ 
Age             & 19.5 kyr        & 20--40 kyr \\ 
Distance        & 3.2 kpc         & 3.4 \\ 
\enddata
\tablecomments{Data are from \citet{Abdo10b} except for G40.5--0.5 distance \citep{yang} and age \citep{Downes80}.}
\end{deluxetable}

\begin{figure*}[t]
\centering
{\includegraphics[width=6.0in]{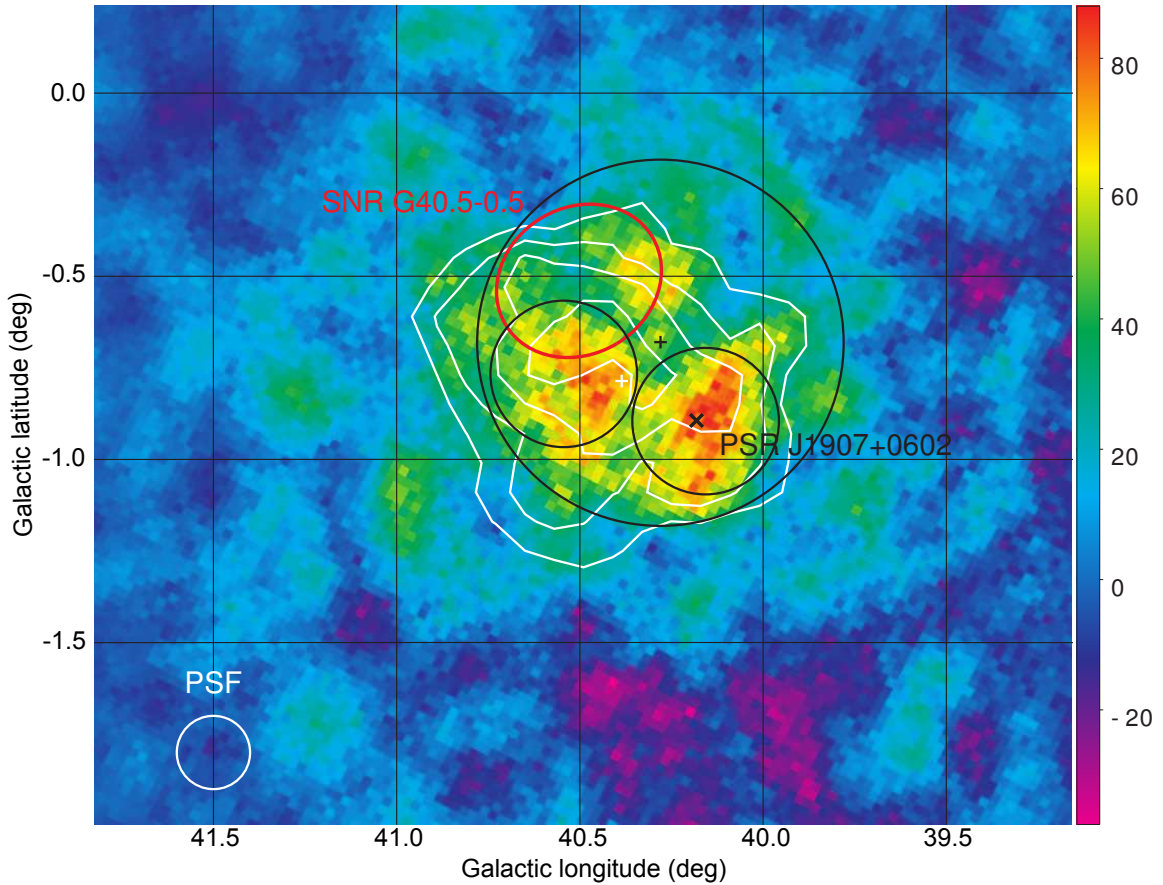}}
\caption{Map of VHE gamma-ray excess counts in the region near MGRO J1908+06.  The color scale shows the excess counts.  The black cross shows the VERITAS best fitted position.  The three black circles show the regions used for spectral analysis.  The white circle shows the size of the VERITAS point spread function (PSF).  The red ellipse shows the extent of SNR G40.5--0.5 in radio continuum emission.  The black `x' shows the location of PSR J1907+0602.  The white cross shows the H.E.S.S. best fitted position and the white contours are from a H.E.S.S.\ excess map at levels of 4, 5, 6, and 7~$\sigma$.  The grids on this and all sky maps in this paper are at intervals of $0.5\arcdeg$.}
\label{fig:gammamap}
\end{figure*}

\section{Observations}

VERITAS is an array of four imaging atmospheric-Cherenkov telescopes (IACTs) located at the Fred Lawrence Whipple Observatory in southern Arizona at an altitude of 1280~m above sea level \citep{veritasstatus}.  Imaging cameras, consisting of 499 photomutiplier tubes located in the focal plane of each telescope, detect Cherenkov light emitted by extensive air showers initiated in the upper atmosphere by gamma rays and cosmic rays. VERITAS has a field of view of 3.5$^{\circ}$ and is sensitive in the range 100 GeV -- 30 TeV.  The system typically operates in `wobble' mode, where the location of the target is offset from the center of the field of view, allowing for simultaneous background measurements \citep{Fomin}. 

The data presented here span six years (2007--2012) and consist of observations taken both before and after the relocation of one of the VERITAS telescopes in 2009 to improve the sensitivity of the array \citep{perkinsmove}, but prior to the upgrade of the VERITAS cameras to high-efficiency photomultiplier tubes during the summer of 2012 \citep{upgrade}.  HESS J1908+063 was initially targeted by VERITAS in 2007--2008, after the announcement of the H.E.S.S.\ detection.  These observations were taken to calibrate the analysis methods used for the VERITAS sky survey \citep{ward, wardsurvey}.  These data were taken both in a ``mini-survey" grid, consisting of a mosaic of exposures covering the source position, and using varying wobble offsets, $0.5\arcdeg$ to $0.7\arcdeg$, from the quoted position for HESS J1908+063.  The north wobble of VERITAS observations of the nearby X-ray binary SS433 \citep{ss433} taken in 2009-2010 provided additional exposure on J1908.  Finally, observations wobbling east and west about a target position approximately halfway between the positions of SS433 and J1908 were taken in 2011--2012.  

The data taken in standard wobble mode, centered on J1908, make up approximately one third of the data set.  North wobbles on the SS433 location make up approximately another one third of the full set.  Finally, the remaining third were taken as east/west wobbles on the location between SS433 and J1908.  The full data set consists of 210 data segments, each lasting, on average, 20 minutes.  After dead-time corrections and removal of periods of bad weather and hardware problems, the total useful exposure is $\sim$62 hours.  The average pointing offset of the telescopes from the centroid of the extended emission is 0.8$^\circ$, and the average zenith angle is 31$^\circ$.

\section{Analysis}
\label{sec:Analysis}

The raw data were calibrated and cleaned, and quality-selection criteria based on the number of photomultiplier tubes contained in the images and the position of the image in the camera were applied.  The shape and orientation of the gamma-ray images were parametrized by their principal moments \citep{hillas}.  The reduction of background cosmic ray events was performed by placing standard selection criteria, optimized using {\em Monte Carlo} simulations and real data from the Crab Nebula, on the shower image parameters.  The following selection criteria, optimized for sources with photon indices of 2.5 or harder, were used in the analysis: `size' or total signal in the telescope images $>$  1000 digital counts ($\sim 200$ photoelectrons), mean scaled width (MSW) between 0.05 and 1.1, and mean scaled length (MSL) between 0.05 and 1.2 \citep{daumwobble,msw}.  For the VHE images and morphology studies presented here, the residual cosmic ray background was estimated using the ring background model \cite[RBM, e.g.,][]{Berge07}.  For the spectral analysis, the reflected-region model was used \citep{refreg}.  Regions with radii of 0.3$^\circ$ around the nearby stars 19 Aquilae and 22 Aquilae as well as the SS433 W2 region \citep[RA, Dec = 287.4, 5.03; ][]{ss433} were excluded from the background estimation.  A region centered on the VHE source with a radius of 0.7$^\circ$ was also excluded from the background analysis.  All of the results presented here were verified using two independent analysis packages.

\begin{figure}[t]
\centering
\includegraphics[width=3.25in]{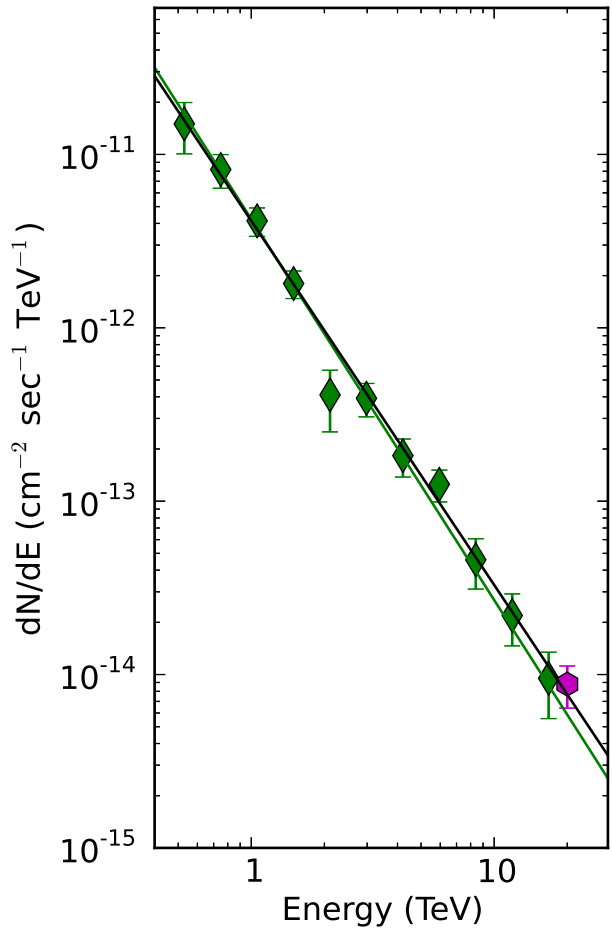} 
\caption{VERITAS spectra for the central $1.0\arcdeg$ (diameter) of J1908.  The green diamonds show the VERITAS spectrum for the whole region ($0.5\arcdeg$ radius region); the green line is a power-law fit to those data.  The black line shows the power-law fit to the HESS J1908+063 data.  The magenta hexagon shows the Milagro measurement.  The VERITAS data are in good agreement with the H.E.S.S.\ and Milagro results.  The image data are available in the electronic edition as the second extension of supplemental FITS file.}
\label{fig:spectra}
\end{figure}

\section{Results} \label{sec:Results}

\subsection{VHE Morphology}
\label{sec:morphology}

Figure~\ref{fig:gammamap} shows an image of the VHE gamma-ray excess counts in the region near MGRO J1908+06.  Each pixel represents an integration over a radius of $0.1\arcdeg$.  There is strong emission near the location of PSR J1907+0602, although the peak of the excess counts is slightly offset to the west.  The TeV emission also extends towards SNR G40.5--0.5.  For comparison, we show the H.E.S.S.\ significance contours for an integration radius of $0.22\arcdeg$ \citep{deOna08}.  The maps agree well in the general localization of the TeV emission, but with a small overall shift discussed further below.

We include a supplemental file containing the gamma-map maps in FITS format.  Each pixel represents an integration over a radius of $0.1\arcdeg$.  The first extension, `SignificanceSkyMap', gives the significance of the gamma-ray detection in each pixel in standard deviations, the second extension, `ExcessSkyMap', gives the gamma-ray excess counts, and is shown in Figure~1.

The TeV image shown in Figure~\ref{fig:gammamap} is a ``correlated excess map.''  The map has a bin size of $0.025\arcdeg$, but the counts in each bin represent a summation of events in which the reconstructed shower arrival direction is within $0.1\arcdeg$ of the center of the bin.  This integration radius is close to the intrinsic angular resolution of the shower reconstruction and our analysis is essentially a standard VERITAS point source analysis.  To quantitatively model the emission morphology, we used an ``uncorrelated excess map,'' where each bin represents a summation only over events with reconstructed shower arrival directions within the bin boundaries.  The measured gamma-ray excesses in different bins are, thus, not correlated.  The uncorrelated excess map used had a bin size of $0.1\arcdeg$.

We first modeled the uncorrelated excess map with a single, symmetric 2D-Gaussian summed with a constant offset.  Examining the residuals, we found that the bulk of the contribution to the $\chi^2$ arose from near the edges of the map where the effective exposure is limited.  We also found that the counts deviated from zero by more than expected from statistical fluctuations in regions away from the source.  To account for these deviations, we included a systematic error of 3.2~counts/bin in our fit.  Restricting the fit to a circular region of radius 1.1$^\circ$ (corresponding to twice the value for $\sigma$ from the initial 2D-Gaussian fit), we found a centroid of (RA, Dec) = ($286.84^\circ\pm0.02^\circ, 6.22^\circ\pm 0.02^\circ)$ and an intrinsic extension of $\sigma_{src} = 0.44^\circ\pm0.02^\circ$.  The fit had a $\chi^2/{\rm DoF} = 983/291$ and the constant offset was $-0.19\pm0.34$.  Using this region of interest, we detect J1908 at a significance level of 14 standard deviations ($\sigma$).  We assign the source name VER J1907+062.  The large values for $\chi^2$/DoF indicate that a simple 2D-Gaussian is a very poor representation  of the true morphology.  Our best fitted position differs from the position reported for HESS J1908+063 by $0.15\arcdeg$.  This is larger than the quoted statistical uncertainties, but a relatively small fraction of the Gaussian width.  The shift may arise due to systematic effects, e.g.\ the fact that the emission is not well described by a symmetric 2D Gaussian or a difference in energy bands between the two measurements.

\subsection{Spectrum}

A spectrum derived from a single source region with a radius of 0.5$^\circ$, to account for the extended nature of the TeV emission and allow direct comparison with the H.E.SS spectrum \citep{Aharonian09}, is shown in Figure~\ref{fig:spectra}.  This region was centered at the source position obtained from the single 2D-Gaussian fit to the $1.1\arcdeg$ region of interest described above.  The diameter of $1.0\arcdeg$ is approximately equal to the full width at half maximum of the fitted Gaussian ($2.35 \sigma_{src} = 1.03^\circ$).  The spectrum was unfolded using VERITAS effective collection areas that were calculated using \emph{Monte Carlo} simulations of extensive air showers passed through the analysis chain with a configuration corresponding to the data-taking conditions and the analysis selection criteria.  The resulting effective areas are used to calculate a differential photon flux.  The differential spectral points were fit with a simple power law of the form:
\begin{equation}
\frac{dN}{dE} = F_0 \times 10^{-12} \times
  \left(\frac{E}{1\mbox{ }TeV}\right)^{-\Gamma_{VHE}}
  \mathrm{photons\mbox{ } cm}^{-2}\mbox{ } \mathrm{s}^{-1} \mbox{ }\mathrm{TeV}^{-1}\;.
\label{eqn:tevspec}
\end{equation}

The spectrum contains 1154 photons after background subtraction, compared to the HESS J1908+063 spectrum which had 689.  The best-fit parameters were $\Gamma_{VHE} = 2.20\pm0.10_{stat}\pm0.20_{sys}$ and $F_0=4.23\pm0.41_{stat}\pm0.85_{sys}$, with $\chi^2/{\rm DoF} = 0.99$.  This result is in good agreement with the results from H.E.S.S.\ and Milagro.  The spectrum is harder and the total flux is lower than measured by ARGO-YBJ, but the difference is significant only at the $2.5\sigma$ level \citep{Bartoli12}.  The Milagro spectrum requires an exponential cutoff between 10 and 40~TeV when the photon index at low energies is fixed to $\Gamma = 2.1$ \citep{Smith10}.  We fitted the VERITAS data with an exponentially cutoff powerlaw and place a 90\% confidence lower limit on the cutoff energy of 17.7~TeV.

We extracted spatially-resolved spectra from two circular regions with $0.2\arcdeg$ radii placed to capture the emission near the PSR and the SNR, but not overlap, see Figure~\ref{fig:gammamap}.  One region contains the pulsar location (`PSR region' below), is centered at RA = 286.968$\arcdeg$, DEC = 6.015$\arcdeg$ (J2000), and has 237 net counts.  The other region is near the boundary of SNR G40.5--0.5 (`SNR region'), is centered at RA = 287.032$\arcdeg$, DEC = 6.418$\arcdeg$, and has 238 net counts.  The angular separation between the centers of the two regions is 0.408$^\circ$.

To search for spectral variations with magnitudes comparable to the systematic error on the absolute spectral index measurement, we performed a differential measurement to minimize the systematic uncertainties.  Specifically, we calculated the ratio of the fluxes of the two spectra in each energy bin and then fitted a power-law to the ratios.  The best fitted photon index for the ratio versus energy is then equal to the difference in photon indices of the two spectra.  Because the response of the array changed when one telescope was relocated in 2009, we calculated separate spectra for the two epochs and then fit the whole set of flux ratios simultaneously.  We find that $\Gamma_{\rm PSR} - \Gamma_{\rm SNR} =  -0.04 \pm 0.20$.  There is no statistically significant evidence for variation in photon index between the two regions.  The spectra of both regions are marginally harder than the spectrum of the $1.0\arcdeg$ diameter region, $\Gamma_{\rm VHE} - \Gamma_{\rm PSR} =  -0.29 \pm 0.12$ and $\Gamma_{\rm VHE} - \Gamma_{\rm SNR} =  -0.20 \pm 0.14$.



\begin{figure*}[t]
\centering
{\includegraphics[width=6.0in]{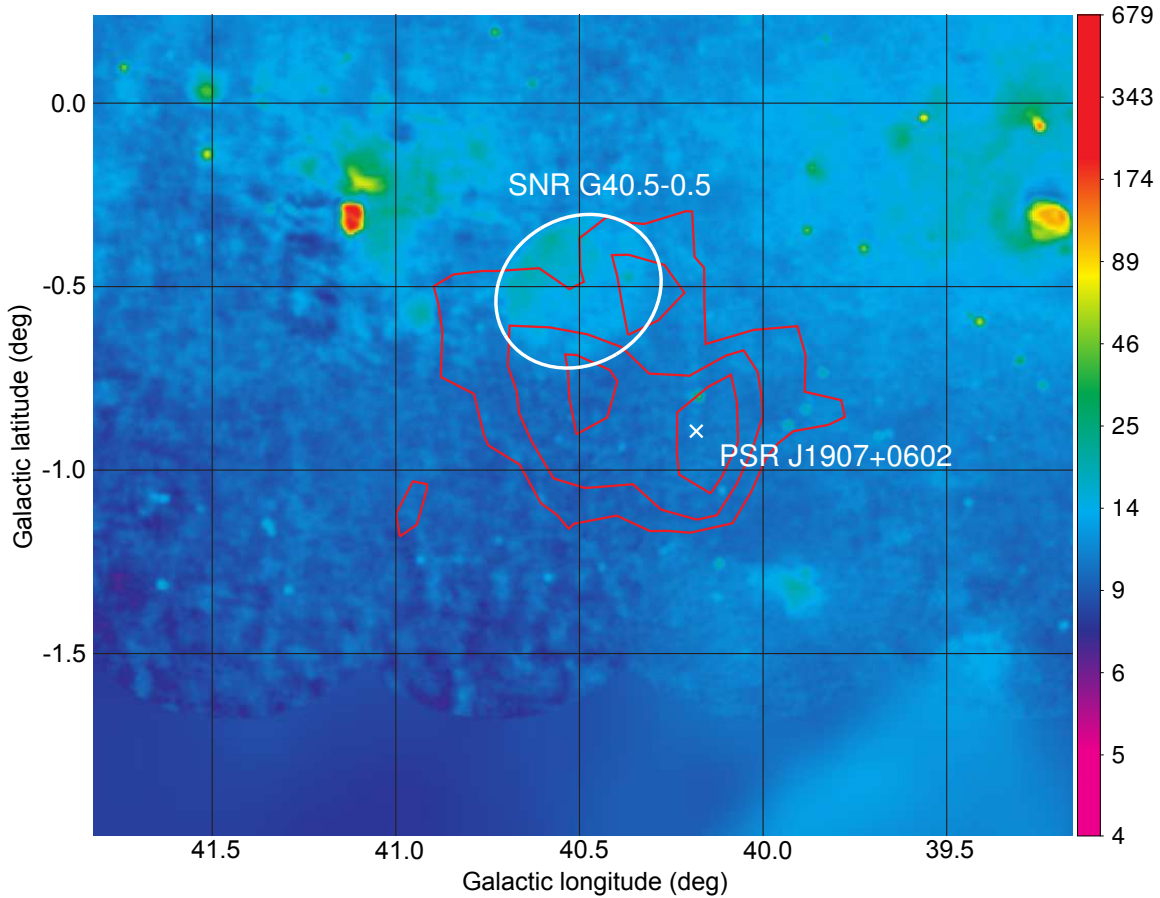}}
\caption{Map of radio continuum intensity at 1420 MHz in the region near MGRO J1908+06.  The color scale shows the brightness temperature in K.  Data are from the VGPS \citep{Stil}.  The red contours show VHE gamma-ray excess counts and are the same as in Figure~\ref{fig:comap}.  The white ellipse shows the extent of SNR G40.5--0.5.  The white `x' shows the location of PSR J1907+0602.}
\label{fig:radiomap}
\end{figure*}

\begin{figure*}[t]
\centering
\includegraphics[width=6.0in]{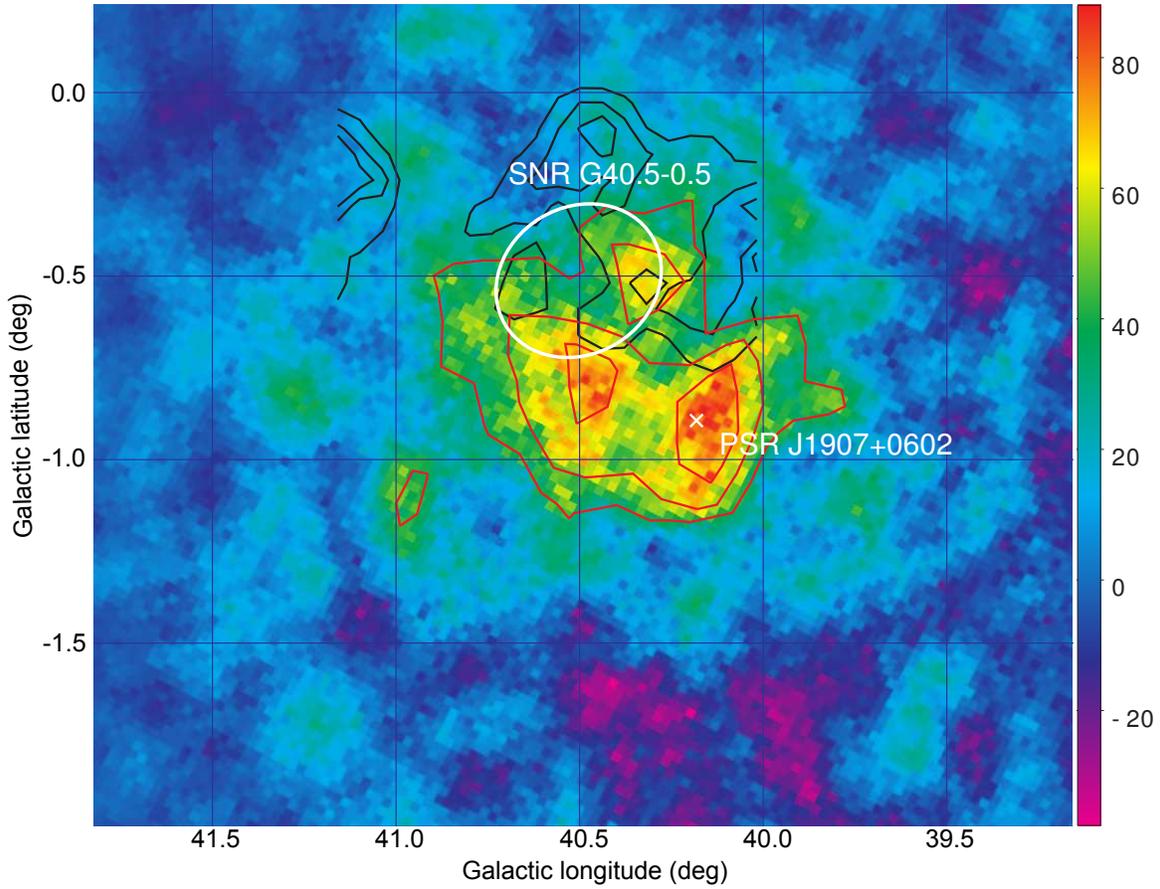} 
\caption{VERITAS map with CO emission contours overlaid.  The black contours show the CO emission and are truncated at the boundaries of the available CO map.  The color scale shows the excess counts and the red contours are at levels of 35, 50, and 65 counts.  These approximately correspond to significance levels of $3\sigma$, $4\sigma$, and $5\sigma$.  The white ellipse shows the extent of SNR G40.5--0.5.  The white `x' shows the location of PSR J1907+0602.}
\label{fig:comap}
\end{figure*}

\begin{deluxetable*}{ccccc}
\tablecaption{Properties of TeV SNRs with ages of 20--60~kyr.
  \label{Table:middle_aged_SNR}}
\tablehead{
Name           & Age        & Luminosity (1--10~TeV) & Radio? & CO? }
\startdata
MGRO J1908+06  & 20--40 kyr & $3\times10^{34}$ erg/s & No     & No  \\
HESS J1731-347 & 27 kyr     & $2\times10^{34}$ erg/s & Yes    & Yes \\
W51C           & 30 kyr     & $2\times10^{33}$ erg/s & Yes    & Yes \\
W28            & 58 kyr     & $2\times10^{33}$ erg/s & Yes    & Yes \\
IC 443         & 1--30 kyr  & $1\times10^{33}$ erg/s & No     & Yes \\
\enddata
\tablecomments{The column ``Radio?'' indicates whether or not there is radio continuum emission coincident with the TeV emission in the SNR.  The column ``CO?'' indicates whether or not there is CO line emission coincident with the TeV emission in the SNR.  References are: HESS J1731-347 \citep{HESSJ1731}, W51C \citep{Aleksic12}, W28 \citep{Aharonian08}, and IC 443 \citep{ver_ic443,Paredes10}. The luminosity of W28 was taken from \citet{Aharonian08}, but for other sources, the luminosities were calculated from the differential fluxes found in the references.}
\end{deluxetable*}

\section{Discussion} \label{sec:Discussion}

The TeV emission from J1908 is strong in the region near PSR J1907+0602 and also extends towards SNR G40.5--0.5.  A key question regarding the nature of the emission is whether it is solely due to a PWN associated with PSR J1907+0602 \citep{Abdo10b} or whether there are additional sources of TeV emission.

We estimate a physical size for J1908 of 50~pc, assuming the pulsar distance of 3.2~kpc and adopting twice the $\sigma = 0.44\arcdeg \pm 0.02\arcdeg$ width of the 2D Gaussian fit as the source diameter.  Comparison with other TeV PWNe shows that J1908 is larger than other TeV PWNe of similar age (see Figure~6 in \citet{Kargaltsev13}\footnote{We note that the size of 70~pc for HESS J1825-137 quoted by \citet{Kargaltsev13} is a factor of $\sim 2$ larger than the size of 36~pc measured in \citet{Aharonian06b}.}).  However, PWN size may depend on environment and interaction with the host SNR, particularly for older PWNe.  Also, J1908 could be made consistent with the size-versus-age relation found for other PWNe if the pulsar is assumed to be a factor of two older than the spin-down age.  This is within the uncertainties associated with spin-down age estimates \citep{Gaensler00} and would also reduce the proper motion to a more typical value.

The 1-10~TeV luminosity of J1908 of $3 \times 10^{34} \rm \, erg \, s^{-1}$ is 1\% of the spin-down power of PSR J1907+0602 of $3 \times 10^{36} \rm \, erg \, s^{-1}$, which gives a TeV gamma-ray efficiency similar to that of other PWNe with pulsars of similar spin-down power \citep{Kargaltsev13}.  X-ray observations reveal hard emission from the location of the pulsar that is possibly extended and could be a compact X-ray PWN or possibly a bow shock \citep{Abdo10b,Pandel12}.  Most, but not all, TeV PWN with similar pulsar spin-down power are detected in X-rays \citep{Kargaltsev13}.  Thus, the gross properties of J1908, except possibly its extent, are compatible with its identification as a TeV PWN powered by PSR J1907+0602.

For J1908, we find that the spectrum measured near the SNR boundary at a projected distance of 23~pc from the pulsar is the same, within uncertainties, as the spectrum measured near the pulsar up to energies of at least 10~TeV, see Figure~\ref{fig:spectra}.  This may present issues with interpretation of the TeV emission as solely powered by PSR J1907+0602.

If PSR J1907+0602 was born at the center of G40.5--0.5, then electrons nearer to the SNR were emitted at earlier times than the electrons near the current pulsar location.  These older populations of electrons would have cooled over the lifetime of the pulsar via synchrotron emission and inverse Compton (IC) scattering.  Assuming an interstellar magnetic field of 3~$\mu$G and that the seed photons for IC scattering are predominately from the cosmic microwave background (CMB), the cooling time for the 63~TeV electrons needed to produce 10~TeV photons is 12~kyr \citep{Kargaltsev07}.  Comparing this cooling time with the pulsar age of $\sim 19.5$~kyr, one would expect significant softening or a fall-off at high energies in the spectrum from the region near the SNR.  Thus, the lack of softening is incompatible with the behavior expected due to electron cooling.  The discrepancy is worse if the characteristic age underestimates the true age, as needed to reduce the pulsar velocity to a more typical value.

Spectral softening is also seen in TeV PWNe containing pulsars without very high proper motions.  The TeV PWN most similar to J1908 is HESS J1825--137 with a physical size of $\sim 36$~pc and a parent pulsar with a characteristic age of 21~kyr, but a proper motion of only $0.023\arcsec$/year \citep{Pavlov08}.  The TeV spectrum of HESS J1825--137 softens with distance from the pulsar, with a photon index of $\Gamma = 1.83 \pm 0.09$ at the pulsar location and $\Gamma = 2.25 \pm 0.06$ at a projected offset of 24~pc \citep{Aharonian06b}.  The change in photon index, $\Delta \Gamma = 0.42 \pm 0.11$ should be compared with that measured above in J1908 at a similar physical separation from the pulsar, $\Delta \Gamma =  -0.04 \pm 0.20$.  Thus, the lack of spectral softening in J1908 far from the pulsar may again argue against interpreting it as a single PWN, even if the pulsar was not born in G40.5--0.5.

A potential counter example is the TeV source TeV J2032+4130 that is a PWN associated with PSR J2032+4127.  VERITAS observations of the source show no spectral evolution across its extent (full width at half maximum) of 11~pc \citep{Aliu14}.  The TeV emission from TeV J2032+4130 appears to be confined to a void apparent in infrared and radio images.  \citet{Aharonian09} note that J1908 lies in a void in molecular line emission integrated over $v_{LSR}$ = 25.3-30.5~km/s corresponding to distances of 1.5--1.8~kpc.  The physical size of J1908 would be compatible with that TeV J2032+4130 at this distance.  However, \citet{Abdo10b} estimate a distance of $3.2 \pm 0.6$~kpc for PSR J1907+0602 based on the pulsar's dispersion measure and place a lower limit on the distance of 3~kpc based on its X-ray flux.  Thus, the distance to PSR J1907+0602 is incompatible with a location in the molecular line emission void unless both pulsar distance estimates are incorrect.

Another source, in addition to the PWN associated with PSR J1907+0602, may contribute to the observed VHE emission.  Fitting the VHE excess map using a model with two 2D-Gaussians provides only marginal improvement, at the $<95$\% confidence level, over the fit using a single 2D-Gaussian.  Thus, we do not claim detection of two separate sources of emission.  However, the observed morphology could be produced by two interacting sources or two sources superimposed along the line of sight.

Some SNRs produce TeV emission via IC interactions of electrons accelerated in the SNR shock or via interactions of accelerated hadrons with surrounding target material such as molecular clouds.  Comparison of the 1420~MHz radio continuum emission from SNR G40.5-0.5 with the TeV emission shows there is little or no radio emission in the regions of strong TeV emission, see Figure~\ref{fig:radiomap}.  Figure \ref{fig:comap} shows the VERITAS excess map with contours of CO line emission integrated over $v_{LSR}$ = 45-65 km/s overlaid (the CO map was kindly provided by \citet{yang}).  The distance to the CO gas at these velocities, 3.4~kpc, is compatible with the distance to the pulsar and the SNR within the uncertainties.  There is CO emission, and thus molecular clouds, at the boundary of the TeV emission, but not coincident with it.  In other middle-aged SNRs with TeV emission, the TeV emission is positionally well correlated with CO line and/or radio continuum emission, see Table~\ref{Table:middle_aged_SNR}.  Thus, it is unlikely that SNR G40.5-0.5 is the main source of the observed TeV emission.

The extended morphology of the TeV emission could be due to interaction of the pulsar wind with molecular clouds or the expanding SNR shell.  A shock formed at such an interaction could reaccelerate energetic particles.  Alternatively, the molecular clouds may provide seed photons for inverse-Compton scattering on high-energy electrons from the pulsar wind.  The dominant seed photons in J1908 are likely microwave photons from the CMB.  IR photons from molecular clouds with typical temperatures of 10--50~K are more energetic than the CMB photons, thus, lower-energy electrons with longer cooling times suffice to produce TeV photons.  The increased radiation energy density, relative to the CMB, near the observed clouds may brighten the TeV emission near the clouds, causing the extended morphology.

In conclusion, J1908 appears to be physically somewhat larger than other TeV PWNe of similar age and the TeV spectrum does not appear to soften with distance from the pulsar, as observed in similar TeV PWNe and expected from electron cooling.  There is no strong TeV emission associated with the radio-bright shell of SNR G40.5-0.5.  Interaction of the pulsar wind with nearby molecular clouds or the SNR shock could explain the large size and lack of spectral softening in J1908.  It is also possible that another PWN, associated with an undetected second pulsar located near the southern edge of SNR G40.5-0.5, could contribute to the VHE emission, but this would require the presence of a second, and as yet undetected, pulsar.


\begin{acknowledgments}

{\it Acknowledgments.} We thank Ji Yang of the Chinese Academy of Sciences' Purple Mountain Observatory for providing the CO maps and the anonymous referee for providing useful comments.  This research is supported by grants from the U.S. Department of Energy Office of Science, the U.S. National Science Foundation and the Smithsonian Institution, by NSERC in Canada, by Science Foundation Ireland (SFI 10/RFP/AST2748) and by STFC in the U.K.  We acknowledge the excellent work of the technical support staff at the Fred Lawrence Whipple Observatory and at the collaborating institutions in the construction and operation of the instrument.  The National Radio Astronomy Observatory is a facility of the National Science Foundation operated under cooperative agreement by Associated Universities, Inc.
\end{acknowledgments}

Facilities: \facility{VERITAS}

\end{document}